\documentclass[reprint,aps,prb]{revtex4-2}
\usepackage{amssymb,amsmath}
\usepackage{graphicx}
\pdfoutput=1

\usepackage{upgreek}

\usepackage{placeins}
\usepackage{bm}
\usepackage{epstopdf}
\usepackage{color}
\usepackage{subfigure}
\usepackage{parskip}
\usepackage[compact]{titlesec}
\usepackage[none]{hyphenat}
\usepackage{hyperref}

\usepackage{esvect}

\hypersetup{
	colorlinks=true,
	linkcolor=blue,
	filecolor=blue,      
	urlcolor=blue,
	citecolor = blue,breaklinks,backref}

\usepackage{cancel}

\titlespacing{\section}{0pt}{1em}{0em}
\titlespacing{\subsection}{0pt}{1em}{0em}

\begin{document}
	
\title{Boundary exceptional point driven chiral  switch of topological edge states}
\author{Bikashkali Midya}
\email{midya@iiserbpr.ac.in}
\author{Barsa Rani Sahu} 
\affiliation{Departmnet of Physical Sciences, Indian Institute of Science Education and Research Berhampur, Odisha 760003, India}

\begin{abstract}
	We theoretically investigate the interplay of topology and localized non-Hermiticity in a finite Su-Schrieffer-Heeger chain characterized by non-Hermitian potentials at its boundaries. By tuning the boundary gain and loss parameters, we exploit the finite-size hybridization of two mid-gap topological edge states to analytically derive the exact coordinates of boundary-induced exceptional points (EPs), $\Gamma_{\rm EP}=\pm i v^{N}/w^{N-1}$, at which these states undergo coalescence. We demonstrate that dynamically encircling these boundary EPs facilitates a robust, chiral edge state-transfer protocol. Depending on the orientation of the encircling loop, this protocol allows a topological edge state localized at one boundary to be either transferred to the opposite edge with perfect fidelity or returned to its starting position through a double switching process.  Importantly, since the non-Hermiticity remains strictly confined to the boundaries, this chiral transport occurs without affecting the energy spectrum of the bulk bands.  These results establish a minimal framework for a large-scale topological chiral switch  that is experimentally verifiable and capable of being controlled through boundary non-Hermitian driving.
\end{abstract}

\maketitle 

\section{Introduction}
The discovery of topological phases of matter has transformed our fundamental understanding of wave localization and transport in quantum and classical systems~\cite{Hasan2010,Ozawa2019}. Central to this paradigm is the principle of bulk-boundary correspondence. This principle  ensures the presence of robust edge-states at the physical boundaries of a system, which is dictated by the non-trivial topological invariants of the bulk energy spectrum. While initially studied within the context of Hermitian quantum systems, topological physics has recently expanded into the non-Hermitian domain to describe open systems subject to gain, loss, or non-reciprocal dissipation \cite{Gong2018,Ghatak2019,Ding2022,Zhang2022,Okuma2023,Lin2023,Coulais2021,Xue2026,Midya2018,Longhi2017}. This integration has revealed distinct phenomena characteristic of non-Hermitian systems, including the non-Hermitian skin effect  \cite{Zhang2022,Okuma2023,Lin2023}, and the appearance of exceptional points (EPs) \cite{Heiss2012,Berry2004,Uzdin2011,Ozdemir2019}, where both eigenfrequencies and eigenstates coalesce.

When the system parameters governing an EP are dynamically encircled in a closed loop, the non-orthogonal characteristics of the coalescing states triggers an asymmetric state-switching process \cite{Uzdin2011,Milburn2015,Hassan2017b}. This mechanism has proven to be effective for chiral mode conversion \cite{Hassan2017,Nasari2022,Doppler2016,Duan2023,Song2021,Chen2021,Yoon2018,Li2020,Liu2021,Ren2022,Xu2016,Zhang2018,Gao2025}, and the generation of entanglement \cite{Khandelwal2024}. It is essential to note that this asymmetric evolution near EPs relies on a non-adiabatic jump phenomenon \cite{Hassan2017b,Zhang2018,Milburn2015}. As the system moves across the complex energy Riemann surfaces, non-adiabatic transitions drive the state vector to jump between instantaneous eigenstates. Since one particular Riemann sheet features a dominant imaginary energy component associated with maximum gain (or minimal loss), any population transitioning to this sheet experiences rapid exponential amplification, which effectively dominates the state vector and directs the wave function into a single target mode determined solely by the chirality of the loop \cite{Hassan2017b,Zhang2018}.

To date, however, the non-Hermitian chiral state-transfer process has been  limited primarily to two-level systems or small-scale configurations (see a recent review \cite{Zhang2026} and the references cited therein). The investigation into scaling this mechanism to large-scale, multi-site lattices is currently in its preliminary stages. A few initial attempts \cite{Zhang2023,Zhang2024,Yu2021,
	Hong2025,Fedrova2020} have relied on global bulk modulations-- such as staggered non-Hermiticity distributed throughout the system-- which unavoidably disrupt the extended bulk energy spectrum. {\it Establishing high-fidelity state transfer process across an extended multi-site lattice through strictly  localized boundary controls, while ensuring the bulk topology remains entirely unperturbed, continues to be an open problem}.

In this study, we theoretically address this discrepancy and discuss a possibility that boundary non-Hermiticity is sufficient to facilitate the chiral transfer of topological edge states while ensuring the bulk spectrum remains entirely undisturbed. Our analysis focuses on a finite Su-Schrieffer-Heeger (SSH) lattice in its topological nontrivial phase, where non-Hermitian potentials are applied exclusively to the first and last sites to represent gain, loss, and detuning at the endpoints. We analytically determine the precise coordinates of the boundary-induced EPs located within the topological gap. We subsequently demonstrate that dynamically encircling these boundary EPs via a closed parameter loop enables a topological edge state, initially localized at one edge, to decouple, traverse the chain, and either re-localize at the opposite edge or return back with perfect fidelity. This process is strictly dependent on the chirality of the encirclement loop. Since the bulk of the chain maintains its Hermitian character, the bulk energy bands and bulk invariants are fully preserved during the operation, thereby confining the non-Hermitian dynamics to the mid-gap edge modes.

A significant theoretical highlight of our model is the derivation of an exact analytical expression for the exceptional point within large-scale systems. Generally, as the lattice size increases, determining the existence of EPs through analytical methods becomes mathematically intractable due to the increasing dimensionality of the system's characteristic polynomial, which typically necessitates a reliance on numerical approximations. In contrast, since our non-Hermitian perturbations are strictly limited to the endpoints, we are able to successfully derive the exact analytical coordinates of the boundary-induced EPs for arbitrary, large system sizes (\(N \gg 1\)). This precise formulation establishes a rigorous mathematical basis for the boundary induced phase transition and offers clear, scalable design principles.

The non-Hermitian transport process described here represents a fundamental shift from the traditional topological transport scheme of the Thouless pumping within a Hermitian Rice-Mele model~\cite{Citro2023,Asboth}. In a standard Thouless pump, the protocol of state transfer from one edge to another necessitates that the initial edge state transiently merges into a bulk state (unless an adiabatic passage is employed~\cite{Longhi2019}), subsequently re-emerging at the opposite boundary through the adiabatic modulation of bulk parameters over a closed cycle~\cite{Asboth}. In contrast, our boundary-driven protocol facilitates edge state chiral pumping across a lattice while ensuring that the edge states remain strictly confined within the topological nontrivial gap throughout the entire dynamical evolution. Since the bulk parameters remain entirely static, the edge states do not transition into bulk. Rather, chiral transport is enabled entirely through the dynamic evolution involving the boundary-controlled EPs, which utilizes non-adiabatic population transfer between topological mid-gap modes, while the bulk ensures the existence of these topological edge states.

The remainder of this paper is organized as follows. In Sec.~\ref{sec:model}, we introduce the boundary-non-Hermitian SSH model and analytically solve for corresponding exceptional points. In Sec.~\ref{sec:dynamics}, we analyze the dynamic encircling protocol and demonstrate the resulting asymmetric, chiral state transfer. Finally, we provide concluding remarks and an outlook in Sec.~\ref{sec:conclusion}.

	\begin{figure}[t!]
	\centering{
		\includegraphics[width=0.5\textwidth]{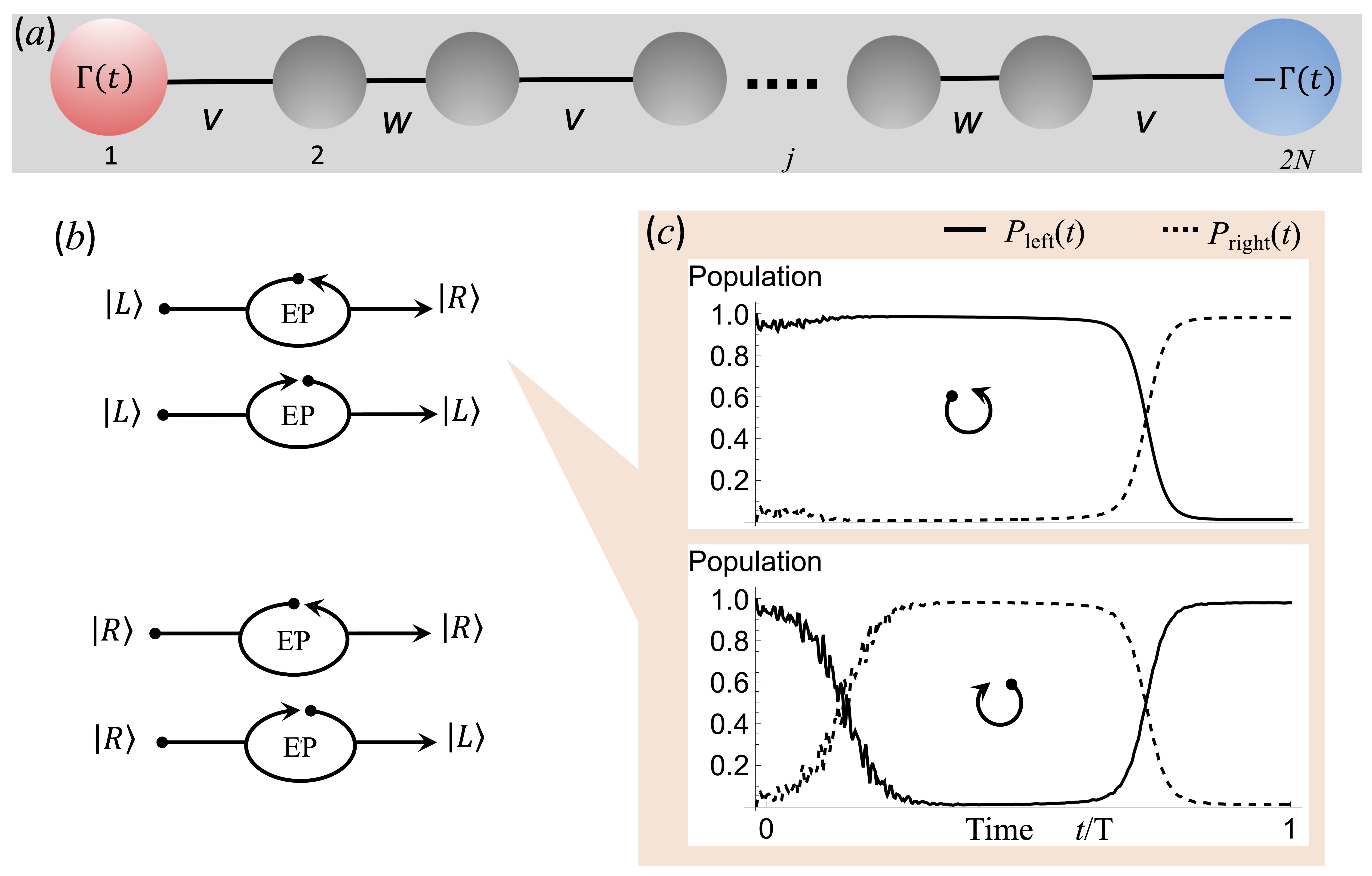} }
	\caption{{\it Scheme of boundary-driven chiral switch of topological edge states  in a finite non-Hermitian lattice.} 	(a) Schematic representation of a finite SSH chain with alternating intra-cell hopping $v$ and inter-cell hopping $w$. Localized non-Hermitian potentials $\Gamma(t)$ and $-\Gamma(t)$ are applied at the leftmost and rightmost sites, respectively, while leaving the bulk entirely Hermitian. 	(b)  Diagram of the asymmetric state switching and returning induced by dynamically encircling the boundary EPs. Depending on the direction of parameter encirclement (chirality of the loop), an initial state localized at the left edge $|L\rangle$ or right edge $|R\rangle$ is either cleanly transferred to the opposite boundary or returns to its initial state.
	(c) An example of time evolution of the dynamical state populations at the left edge ($P_{\text{left}}$, solid lines) and right edge ($P_{\text{right}}$, dashed lines) over a single  cycle for a lattice of $2N=16$ sites, with the population initially  localized at the leftmost site $j=1$. A $\circlearrowleft$ or  $\circlearrowright$ loop dictates a high-fidelity chiral state transfer or robust state return. Complete  dynamics is discussed below in Fig.~\ref{fig-3}. } \label{fig-1}
\end{figure}

\section{\label{sec:model} Boundary non-Hermitian SSH model and its analytical EPs}
We consider a  finite SSH lattice consisting of \(N\) unit cells, where each cell contains a two-site basis corresponding to odd and even  sublattices, yielding a total of \(2N\) lattice sites~\cite{Asboth} (Fig.~\ref{fig-1}(a)). The bulk of the chain is strictly Hermitian, characterized by alternating, static intra-cell hopping amplitude \(v\) and inter-cell hopping amplitude \(w\). To isolate the non-Hermitian effects entirely at the boundaries, we introduce time-dependent complex potentials, denoted as \(\Gamma(t)\) and \(-\Gamma(t)\),  located at the left $(j=1)$ and right $(j=2N)$ endpoints of the lattice, respectively. The total tight-binding Hamiltonian \(\mathcal{H}\) of the system can be written as:
\(\mathcal{H}(t)=\mathcal{H}_{\text{bulk}}+\mathcal{H}_{\text{boundary}}(t)\)
where the Hermitian bulk Hamiltonian takes the standard form:
\begin{equation}
	\mathcal{H}_{\rm bulk}= \sum_{j=1}^{2N-1} \kappa_j (a_j^\dag a_{j+1} + a_{j+1}^\dag a_{j}),
\end{equation}
The operators $a_{j}^\dagger$ ($a_{j}$) denote the bosonic creation (annihilation) operators at site $j$, and the alternating hopping amplitudes are defined as
\begin{equation}
	\kappa_j = \begin{cases} 
		v & \text{if } j \text{ is odd}, \\ 
		w & \text{if } j \text{ is even}. 
	\end{cases}
\end{equation}
The non-Hermitian boundary term is given by 
\begin{equation}
	\mathcal{H}_{\rm boundary}(t) = \Gamma(t) \left( a_1^\dagger a_1 - a_{2N}^\dagger a_{2N} \right),
\end{equation}
where the time-dependent complex potential is defined as
\begin{equation}
	\Gamma(t) = \delta(t) + i \upgamma(t).
\end{equation}
Here, $\delta(t)$ controls the dynamic local energy detuning relative to the bulk mid-gap energy (set here to zero), and $\upgamma(t)$ represents the gain or loss rate required to encircle the EPs. We choose weak non-Hermiticity regime $|\upgamma(t)|\ll |v-w|$, so that boundary state do not mix with bulk continuum.

\begin{figure}[t!]
	\centering{
		\includegraphics[width=0.5\textwidth]{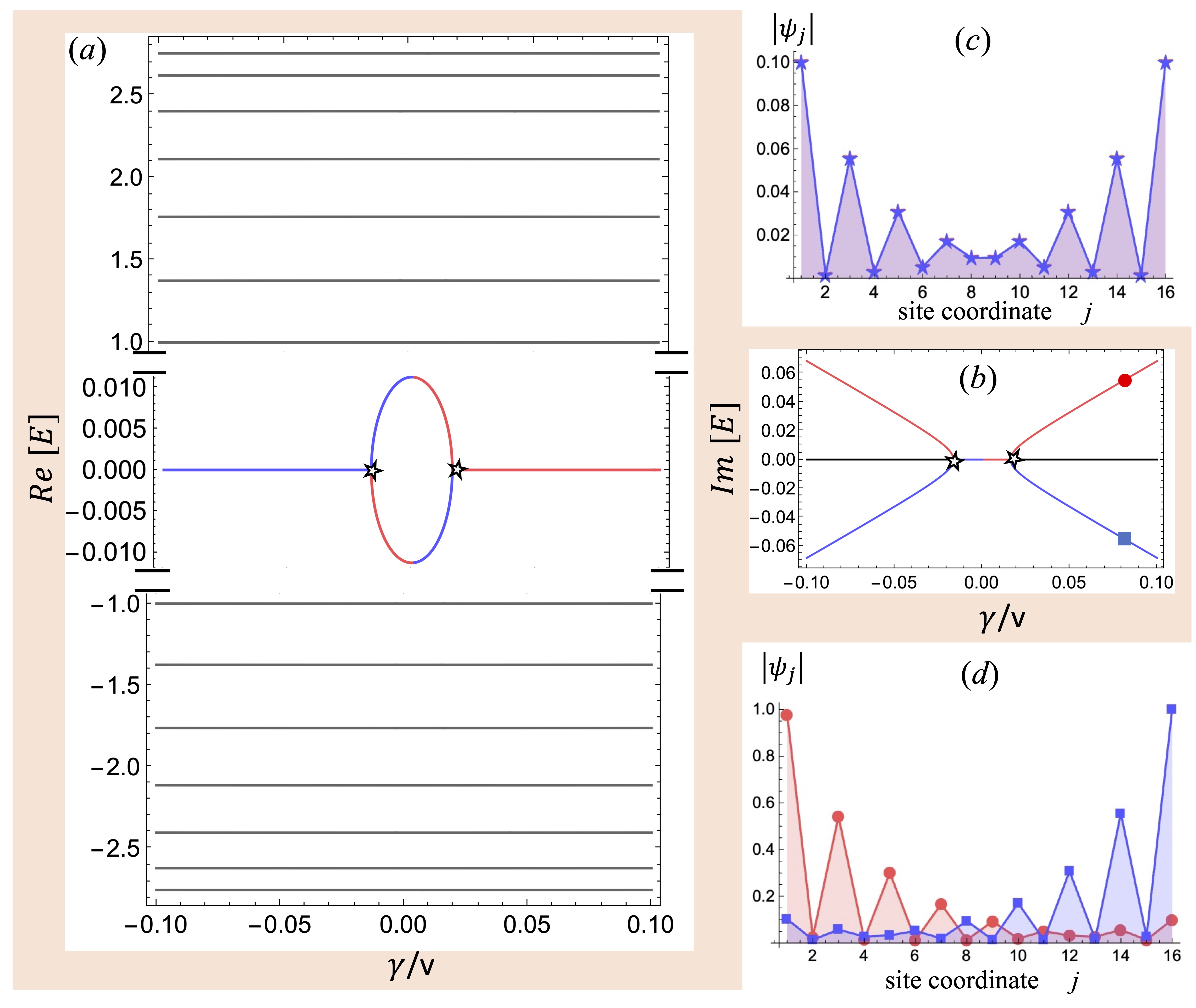} }
	\caption{{\it  Energy spectrum, boundary EPs, and spatial field profiles of a finite non-Hermitian SSH lattice} with $2N=16, w/v=1.8, \delta=0$. (a)~Real part $\text{Re}[E]$, and (b)~imaginary part $\text{Im}[E]$ of the energy spectrum, obtained through exact numerical diagonalization, are plotted as a function of the dimensionless non-Hermitian strength $\upgamma/v$. The bulk bands remain entirely real, and completely unperturbed by the localized non-Hermitian perturbations. Within the mid-gap region, the topological edge states undergo a localized phase transition. The two distinct boundary-induced EPs, where both eigenvalues and eigenvectors coalesce, are explicitly marked by $\star$.	The edge states exhibit purely real energies in the $\mathcal{PT}$-unbroken regime ($\vert\upgamma\vert < \vert\upgamma_{\text{EP}}\vert$) and bifurcate into a complex conjugate pair (red and blue branches) in the $\mathcal{PT}$-broken regime ($\vert\upgamma\vert > \vert\upgamma_{\text{EP}}\vert$).
		(c) Nonnormalized edge state probability distribution $|\psi_j|$ across the lattice coordinate $j=1,2,\cdots, 16$. The top panel illustrates coalescence of hybridized edge-state wavefunctions at the exceptional points ($\upgamma_{\text{EP}}/v \approx \pm 0.016$), while (d) shows the sharp, asymmetric single-edge localization of the red (circle) and blue (square) modes in the broken phase ($\upgamma/v = 0.08$). The energy plotted here is in the unit of coupling $v$. } \label{fig-2}
\end{figure}

For a finite lattice in the parameter regime $v<w$, $\mathcal{H}_{\rm bulk}$ is topologically nontrivial and hosts two localized midgap edge states, whereas the system is topologically trivial and possesses only extended bulk states when $v \ge w$~\cite{Asboth}. {\it Throughout this work, we focus exclusively on the topologically nontrivial bulk configuration~($v<w$)}. Within this regime, the finite size of the lattice prevents the two midgap edge states from being perfectly isolated to a single edge. Instead, their wavefunctions penetrate the bulk and exponentially overlap, resulting in a finite-size hybridization that lifts their zero-energy degeneracy and splits their individual energies into a distinct pair denoted as $E_+$ and $E_-$~[Fig~\ref{fig-2}(a)].

When the boundary term $\mathcal{H}_{\rm boundary}$ is introduced, this hybridization can be exactly counterbalanced by the gain and loss parameters, driving the two edge states to coalesce at an EP where their split energies converge to a single degenerate value $E_+ = E_- = 0$. To determine the exact location of these boundary-induced EPs, we express the full system Hamiltonian in a $2N \times 2N$ matrix form with respect to the site basis~$\lvert\psi\rangle=\sum_{j=1}^{2N}\psi_j \lvert j\rangle$, where $|\psi_j|^2$ represents the population probability at site~$j$:
\begin{equation}
	\mathcal{H} = \begin{pmatrix}
		\Gamma & v & 0 & 0 & \dots & 0 & 0 \\ 
		v & 0 & w & 0 & \dots & 0 & 0 \\ 
		0 & w & 0 & v & \dots & 0 & 0 \\ 
		0 & 0 & v & 0 & \dots & 0 & 0 \\ 
		\vdots & \vdots & \vdots & \vdots & \ddots & \vdots & \vdots \\ 
		0 & 0 & 0 & 0 & \dots & 0 & v \\ 
		0 & 0 & 0 & 0 & \dots & v & -\Gamma
	\end{pmatrix}.
\end{equation}
The full eigenvalue problem $\mathcal{H}|\psi\rangle = E|\psi\rangle$ is equivalent to the following set of $2N$ coupled difference equations:
\begin{equation}
	\begin{aligned}
		\Gamma\psi_1 + v \psi_2 &= E \psi_1, \\
		\kappa_{j-1}\psi_{j-1} + \kappa_j \psi_{j+1} &= E \psi_j, \quad (j = 2, \dots, 2N-1), \\
		v\psi_{2N-1} - \Gamma\psi_{2N} &= E \psi_{2N}.
	\end{aligned}
\end{equation}
At an EP $(E=0)$, evaluating the spatial recurrence from the left boundary yields the final site amplitude:
\begin{equation}
	\psi_{2N} = (-1)^N \Gamma \frac{w^{N-1}}{v^{N}} \psi_1.
\end{equation}
Similarly, propagating the recurrence relations from the right boundary condition dictates that:
\begin{equation}
	\psi_1 = (-1)^{N+1} \Gamma \frac{w^{N-1}}{v^{N}} \psi_{2N}.
	\end{equation}
		Demanding the self-consistency of these two boundary-matching equations directly yields the exact analytical coordinates for the EPs:
		\begin{equation}
			\Gamma_{\rm EP} = \pm i \frac{v^{N}}{w^{N-1}}.\label{eq:EP}
		\end{equation}
The system possesses two conjugate EPs  at $(\delta_{\rm EP},\upgamma_{\rm EP})=(0, \pm v^{N}/w^{N-1})$. In the limit of $N=1$ unit cell, the EP coincides with that of a prototype two-state non-hermitian system studied previously~\cite{Hassan2017b}. This analytical form of EP valid for arbitrarily large lattices is one of the core results of our study.  A detailed numerical validation of this spectral behavior is illustrated in Fig.~\ref{fig-2}. 

The occurrence of above EPs can also be intuitively understood by projecting the full lattice problem onto a reduced two-dimensional subspace spanned by the decoupled boundary states \(\{\lvert L\rangle, \lvert R\rangle\}\) pinned to the leftmost and rightmost edges. In the topologically non-trivial regime (\(v < w\)), the two hybridized eigenstates can be expressed as symmetric and antisymmetric combinations of these decoupled basis states:
\begin{equation}|\psi _{\pm }\rangle \simeq \frac{1}{\sqrt{2}}\left(|L\rangle \pm |R\rangle \right),\end{equation}
where 
\begin{equation}\begin{array}{ll}
	|L\rangle \simeq\left[1,0, \left(-\frac{v}{w}\right),0, \cdots,  \left(-\frac{v}{w}\right)^{N-1},0\right]^T,\\\\
	|R\rangle \simeq\left[\begin{matrix}0,\left(-\frac{v}{w}\right)^{N-1}, 0, \cdots, \left(-\frac{v}{w}\right), 0, 1\end{matrix}\right]^T.
	\end{array}
	\end{equation}
Projecting the full system Hamiltonian onto this localized boundary basis \(\{\lvert L\rangle, \lvert R\rangle\}\) yields the effective two-level non-Hermitian Hamiltonian:
\begin{equation} \mathcal{H}_{\text{eff}} \simeq \begin{pmatrix} \delta + i\upgamma & \kappa_{\text{edge}} \\ \kappa_{\text{edge}} & -\delta - i\upgamma \end{pmatrix}, \end{equation} where the off-diagonal element 
\begin{equation} \kappa_{\text{edge}} = \langle L|\mathcal{H}_{\text{bulk}}|R\rangle=(-1)^{N-1}\frac{v^{N}}{w^{N-1}},
\end{equation}
represents the passive, bulk-mediated effective coupling between the two decoupled edge states. The complex eigenvalues of this effective boundary matrix take the form:
	\begin{equation}E_{\text{eff}, \pm}=\pm \sqrt{\left(\delta+i\upgamma \right)^{2}+\kappa _{\rm edge}^{2}}.\end{equation}
Remarkably, this simple two-state approximation yields the exact same phase transition threshold, \(\upgamma_{\rm{EP}}=\pm|\kappa_{\rm edge}|=\pm v^N/w^{N-1}\), derived, in the form Eq.~\ref{eq:EP}, from the full multi-site recurrence relations.  This perfect agreement confirms that the state coalescence is driven strictly by a localized competition between on-site non-Hermitian parameters and passive, finite-size bulk mediated tunneling.

 In the absence of boundary real potential detuning $\delta=0$, the system is parity-time $(\mathcal{PT})$ symmetric \cite{Bender}.  However, the symmetry is spontaneously broken at $\upgamma=\upgamma_{\rm EP}$.  One has the \(\mathcal{PT}\)-unbroken phase (\(\vert{}\upgamma\vert{} < \vert{}\upgamma_{\text{EP}}\vert{}\)): when the boundary dissipation rate \(\upgamma \) is weaker than the bulk-mediated edge coupling \(\kappa _{\text{edge}}\), coherent tunneling between edges dominates, and all the energy spectrum is real. As the gain/loss parameter is increased, it drives the split edge energies \(E_{+}\) and \(E_{-}\) to contract along the real axis. When the on-site non-Hermitian strength perfectly counterbalances the bulk-mediated hybridization (\(\upgamma =|\kappa_{\text{edge}}|\)), the initial energy splitting entirely vanishes, forcing both the eigenvalues and their corresponding eigenvectors to coalesce at \(E_+ = E_- = 0\). The edge spectrum become complex beyond $|\upgamma|>|\upgamma_{\rm EP}|$, (numerical example is given in Fig.~\ref{fig-2}).

\begin{figure*}[t!]
	\centering{
		\includegraphics[width=0.9\textwidth]{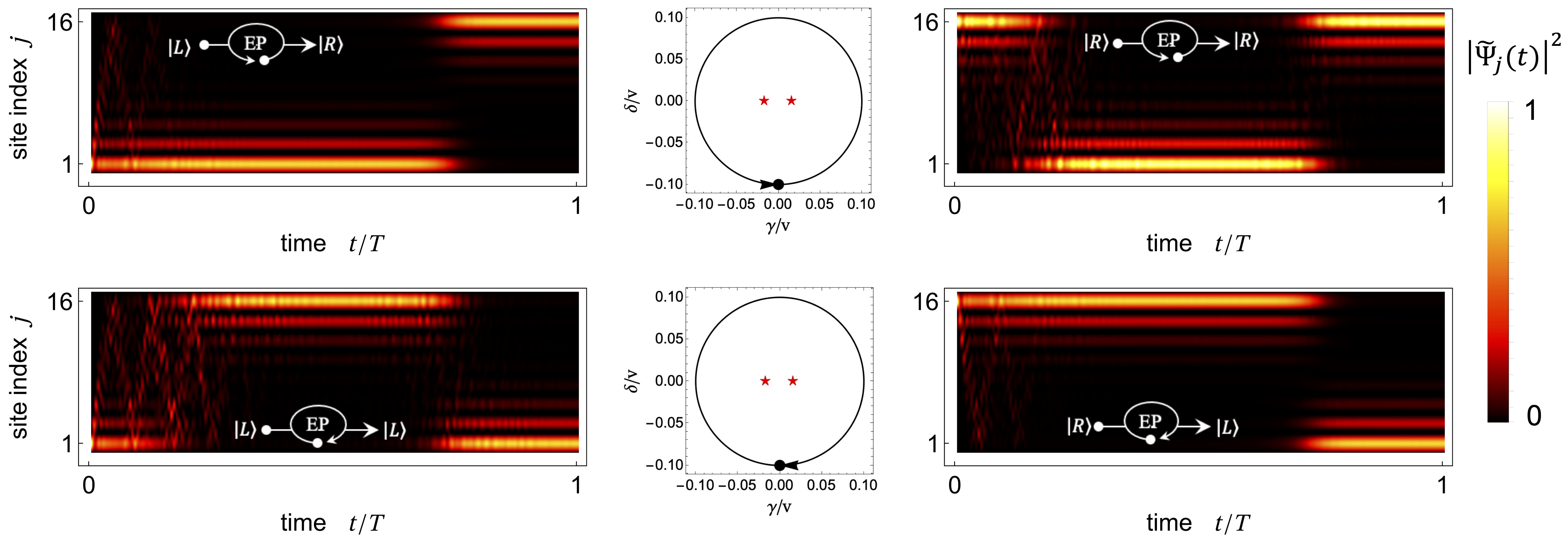} }
	\caption{ {\it Time-resolved dynamical chiral state transfer and return via EPs encirclement}. 	Center column: closed parameter trajectories in the $(\upgamma/v, \delta/v)$ plane enclosing the two boundary-induced EPs (marked by red stars). The black dots denote the starting points of the dynamic cycle, and the arrows indicate the loop chirality: $\circlearrowleft$ (top) and $\circlearrowright$ (bottom).	Left column: the dynamic evolution of the initial single-site excitation situated at the left edge $j=1$, expressed as a function of the normalized time $t/T$. Under the $\circlearrowright$ protocol (top left), the mode smoothly transits across the bulk to completely re-localize at the opposite right boundary ($j=16$). Under the $\circlearrowleft$ protocol (bottom left), the population returns and remains locked to the initial left boundary.  Right column: corresponding state evolution initialized in the single site $j=16$ at the right. Under the $\circlearrowleft$ protocol (top right), the mode returns at the right edge, whereas under the $\circlearrowright$ protocol (bottom right), it transits across the chain to localize at the left edge ($j=1$). Here, we have chosen $2N=16, w=1.8v, T=200v^{-1}, \theta_0=\pi, r=0.1v$, and $v=1$. Corresponding $\delta_{\rm EP}=0, \upgamma_{\rm EP}\simeq\pm 0.016 v$. } \label{fig-3}
\end{figure*}

\section{\label{sec:dynamics}Chiral edge state transfer by dynamical encircling EPs}
To implement the chiral transport protocol, we vary parameters dynamically in one time cycle $0\le t\le T$, defining a closed loop in the complex $\Gamma=(\delta, \upgamma)$ plane. Unlike conventional non-Hermitian protocols that encircle a single isolated EP, our loop parameterization encircles both EPs (\(0, -\kappa_{\text{edge}}\)) and (\(0, \kappa_{\text{edge}}\)) simultaneously. This dual-EP encirclement strategy is fundamentally dictated by our core objective: {\it achieving a complete, high-fidelity physical state transfer from one far edge of the extended lattice to the other}. In a finite chain, a single-EP loop typically operates within a narrow parameter regime where the boundary states remain continuously hybridized and partially delocalized across both ends of the lattice. While a single-EP encirclement is verified to induce state switching, it fails to execute clean cross-chain transport. To pump a state from one edge to the other, the initial and final configurations of the dynamic cycle must rest in a Hermitian regime $(\upgamma=0)$, and  where the edge states are completely decoupled from each other and strictly localized on their respective endpoints. This decoupling is achieved by applying a large real energy detuning at the start and end of the cycle (\(\vert\delta(0)\vert=\vert\delta(T)\vert \gg |\kappa_{\text{edge}}|\)). When the detuning dominates over the bulk-mediated coupling, the finite-size hybridization is entirely suppressed, forcing the eigenstates to lock purely onto respective lattice edges. Consequently, the chosen closed  loop  over a total cycle duration \(T\) is as follows:
\begin{equation}
	\Gamma(t)=\Gamma_0 + r e^{i (s \theta(t)+\theta_0)}
\end{equation}
where $\theta(t)=2\pi t/T$, $\theta_0$ is the initial phase, $\Gamma_0=(\delta_0, \upgamma_0)$ represents the center of the loop, and \(r\) denotes the loop radius. To ensure the complete encirclement of both boundary EPs, we center the loop at the origin (\(\delta_0 = 0, \upgamma_0 = 0\)) and select a large radius satisfying \(r > |\kappa_{\text{edge}}|\). The direction of the parameter sweep—and consequently the chirality of the pumping protocol—is dictated by the sign of $s$, where \(s =-1 \) corresponds to a counter-clockwise ($\circlearrowleft$) trajectory and \(s =1\) corresponds to a clockwise ($\circlearrowright$) trajectory. In all the numerical examples discussed below, we have chosen the encircling loop starting $(t=0)$ and ending  $(t=T)$ points such that the  Hamiltonian $\mathcal{H}(0)\equiv\mathcal{H}(T)$ is hermitian.

\begin{figure*}[ht!]
	\centering{
		\includegraphics[width=0.8\textwidth]{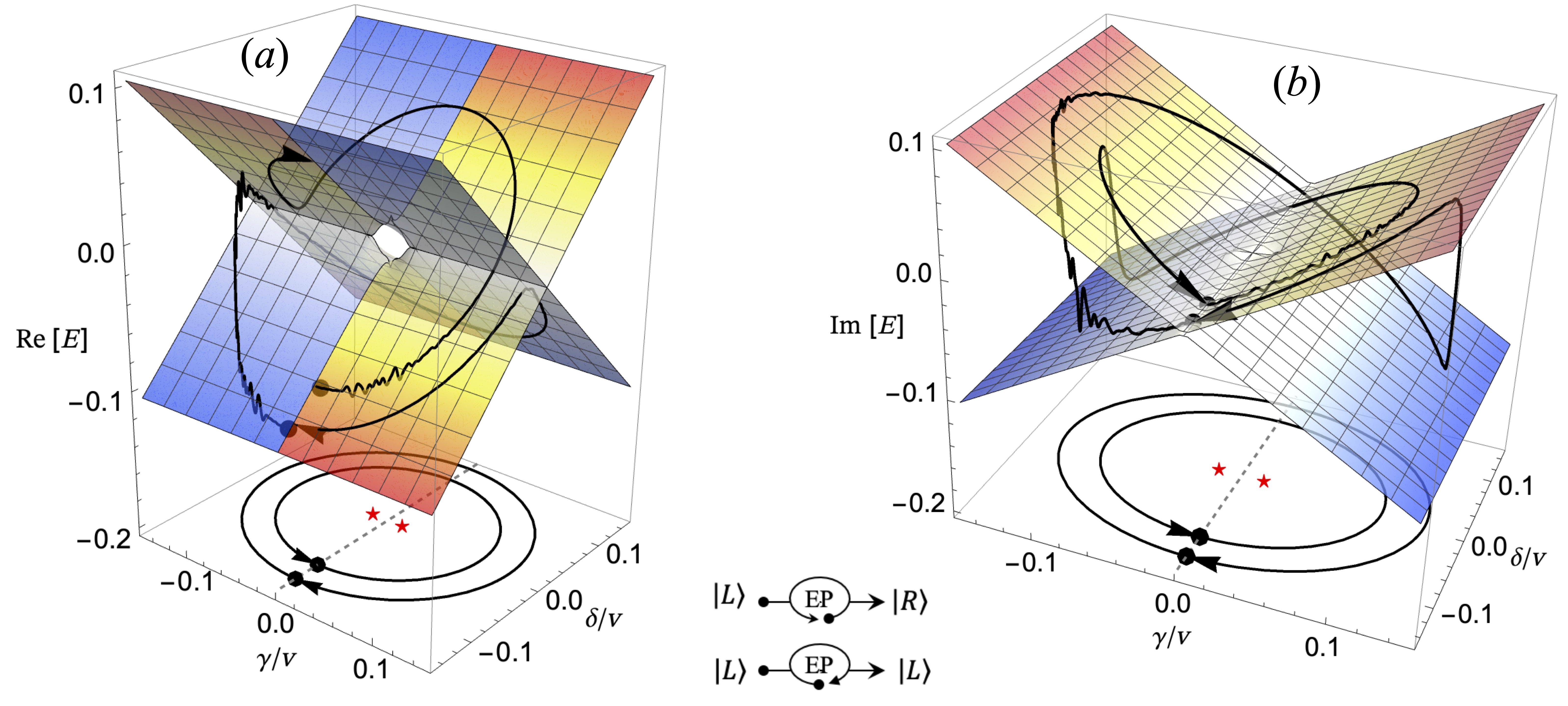} }
	\caption{{\it Nonadiabatic dynamical energy trajectories on edge-state eigenenergy Riemann surfaces}. (a) Real part, and (b) imaginary part of the instantaneous midgap energy eigenvalues plotted as a function of $\upgamma/v$ and $\delta/v$. The red (blue) surface represents gain (loss) sheet with positive (negative ) imaginary part; bulk eigenenergies are omitted for clarity. The intersecting, non-Hermitian Riemann sheets illustrate the characteristic branch-point singularity dictated by the boundary-induced EPs. The projection on the base plane explicitly marks the positions of the two coalescent exceptional points (red stars). The solid black curves on the Riemann surfaces track the dynamic evolution of the energy $E(t)$ corresponding to state $\lvert\Psi(t)\rangle$ shown earlier in the left column of Fig.~\ref{fig-3}, for $\circlearrowleft$ and $\circlearrowright$ encircling starting from excitation at the leftmost $(j=1)$ site. Clearly, $\circlearrowleft$ encircling facilitates state swapping (see ${\rm Re}[E]$) by a single non-adiabatic jump (see ${\rm Im}[E]$), while $\circlearrowright$ encircling allows the state to return (${\rm Re}[E]$) to its origin after two non-adiabatic jump (see ${\rm Im}[E]$). Here, we have chosen $2N=16, w=1.8v, T=200v^{-1}, \theta_0=\pi, r_\circlearrowleft=0.1v$, $r_\circlearrowright=0.12v$, and $|\Psi(0)\rangle=[1,0,\cdots]^T$ for the computation of $|\Psi(t)\rangle$ and $E(t)$.} \label{fig-4}
\end{figure*}

\subsection{Dynamical chiral state transfer}
The dynamical evolution of an an initial excitation \(\lvert\Psi(0)\rangle\) is governed by the time-dependent Schr\"odinger equation:~\(i\frac{\partial }{\partial t}|{}\Psi (t)\rangle=\mathcal{H}(t)|{}\Psi (t)\rangle \).  Because the time-dependent boundary parameters cause the Hamiltonian matrices at distinct time steps to be in general non-commuting (\([\mathcal{H}(t_1), \mathcal{H}(t_2)] \neq 0\)), the formal solution to the state vector evolution~$|\Psi (t)\rangle=\sum_{j=1}^{2N}\Psi_j(t)\lvert j\rangle$ at an arbitrary time \(t\) within the cycle is given by:
\begin{equation}|\Psi (t)\rangle=\mathcal{T}\left\{\exp \left[-i\int _{0}^{t}\mathcal{H}(t^{\prime })dt^{\prime }\right]\right\}|\Psi (0)\rangle
\end{equation}
where $\mathcal{T}$ is the time-ordering operator. In order to demonstrate, we consider a lattice with $2N=16$ sites and numerically obtain $\lvert\Psi(t)\rangle$ for both $\circlearrowleft$ and $\circlearrowright$ encirclement starting from a single site excitation either at left or right edge i.e. $\lvert\Psi(0)\rangle=[1,0,0,\cdots0]^T$ or $\lvert\Psi(0)\rangle=[0,0,\cdots,0, 1]^T$. The corresponding  dynamical evolution of normalized site populations~$|\widetilde{\Psi}_j(t)|^2=|\Psi_j(t)|^2/\sum_j |\Psi_j(t)|^2$ is shown and discussed in Fig.~\ref{fig-3}. For a given loop chirality, the system is unidirectionally  driven into the same target edge state: right edge for $\circlearrowleft$ and left edge for $\circlearrowright$ regardless of the initial state preparation.  While the passive bulk of the lattice remains unperturbed, but crucially it provides topological stability to the existence of robust edge state. These results demonstrate the non-Hermitian chiral dynamical signature of the EPs in a topological lattice. 

Due to the sublattice polarization of the SSH edge states~\cite{Asboth}, the left edge mode is confined mainly to the odd site indices and the right edge mode occupies the even site indices. We therefore define instantaneous population at the left-, and right-edge as:
\begin{equation}
	\small	P_{\text{left}}=\frac{\sum_{j=1}^{N} |\Psi_{2j-1}(t)|^2}{\sum_{j=1}^{2N} |\Psi_j(t)|^2}, ~ P_{\text{right}}=\frac{\sum_{j=1}^{N} |\Psi_{2j}(t)|^2}{\sum_{j=1}^{2N} |\Psi_j(t)|^2}, 
\end{equation}
respectively. Because the bulk parameters remain static and the bulk topological gap never closes, the extended bulk bands remain unpopulated. Consequently, this normalized sublattice intensity tracking provides a direct, experimentally accessible measure of the cross-chain transfer and double-switch return dynamics. Numerical example of the instantaneous left and right edge population for initial excitation at site $j=1$ is shown in Fig.~\ref{fig-1}(c), which corroborates with  the corresponding full dynamical state evolution given in Fig.~\ref{fig-3} (left column).

  	\subsection{Non-adiabatic dynamical energy evolution on the Riemann surfaces}
  	To directly trace the non-adiabatic jump dynamics during the encirclement cycle, as observed in Fig.~\ref{fig-3}, we track the instantaneous expectation value of the Hamiltonian, defined as the dynamical energy:
  	\begin{equation}
  		E(t) = \frac{\langle \Psi(t) \rvert \mathcal{H}(t) \lvert \Psi(t) \rangle}{\langle \Psi(t) \vert \Psi(t) \rangle}.
  	\end{equation} 
  	Because the system is non-Hermitian, $E(t)$ is a complex quantity whose real and imaginary components can be dynamically mapped as a continuous trajectory superimposed onto the instantaneous edge state eigenenergy Riemann surfaces except when non-adibatic transition occurs [Fig.~\ref{fig-4}]. These Riemann surfaces, as a function of $\upgamma/v$ and  $\delta/v$, are characterized by a non-trivial branch-cut topology that splits into distinct gain ($\text{Im}[E_{\rm inst}] > 0$) and loss ($\text{Im}[E_{\rm inst}] < 0$) sheets, where $E_{\rm inst}$ is the quasi-static instantaneous edgestate eigenvalues. As the closed parameter loop winds around the EPs, the time-dependent variation of the boundary potentials induces strong non-adiabatic coupling between two edge states~\cite{Milburn2015,Zhang2018}. This coupling causes the dynamical energy $E(t)$ to deviate from the pure adiabatic instantaneous energy surfaces, manifested as characteristic non-adiabatic jump actively forces the state vector onto the dominant gain sheet  [Fig.~\ref{fig-4}]. The resulting exponential amplification of this specific Riemann branch rapidly dominates the state vector, efficiently filtering out the decaying mode component. The final state of the system depends strictly on the chirality of the encirclement loop rather than the initial state preparation~\cite{Zhang2018}. When the loop is traversed in a counter-clockwise ($\circlearrowleft$) direction, the dynamical expectation value $E(t)$ tracks a path where the left boundary mode $\lvert L\rangle$ transfers its population entirely to the right boundary mode $\lvert R\rangle$. Conversely, reversing the sweep to a clockwise ($\circlearrowright$) direction triggers a multi-stage double-switch trajectory; the dynamical energy transits toward the right edge before the full-system sheet topology forces a second non-adiabatic jump that pumps the population back to its initial left boundary [see left column of Fig.~\ref{fig-3}, and Fig.~\ref{fig-4}]. Opposite behavior is observed when an excitation at right boundary is allowed to evolve (shown in right column of  Fig.~\ref{fig-3}, but the corresponding evolution on the Riemann manifold is not shown). Throughout this complex sheet evolution, the extended bulk bands remain widely separated by a robust topological gap and stay entirely passive, confirming that the dynamic state manipulation is governed exclusively by the non-trivial topology of the boundary-localized edge sheets.

\begin{figure}[t!]
	\centering{
		\includegraphics[width=0.45\textwidth]{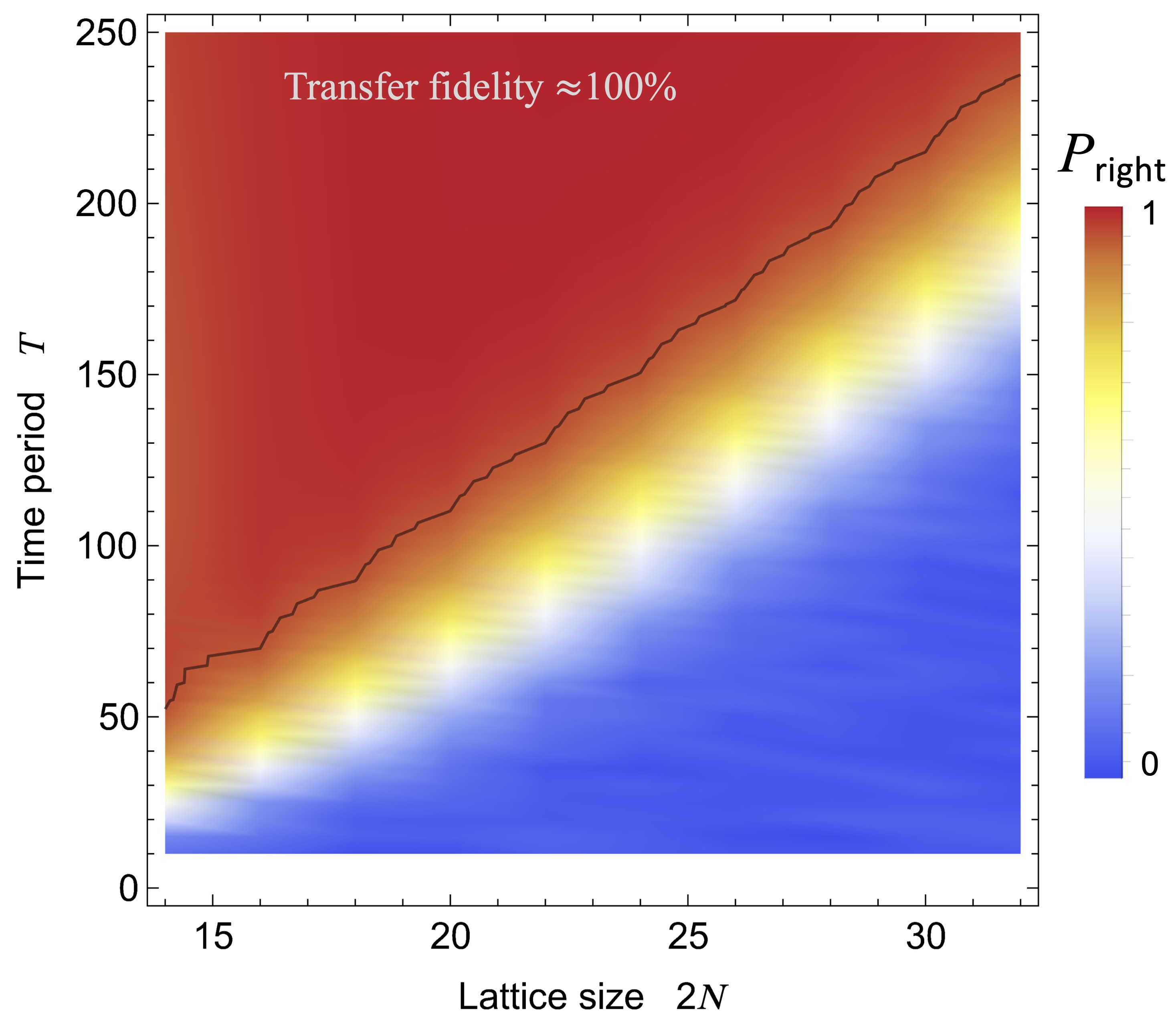} }
	\caption{Pseudo color  plot of the left-to-right topological edge state transfer fidelity as a function of the encircling time period $T$ and the lattice size $2N$. The color scale quantifies $P_{\rm right}(t=T)$, the transfer fidelity, ranging from $0$ (low fidelity, blue) to $1$ (perfect fidelity, red), calculated for a fixed $\circlearrowleft$ loop radius  $r=0.1v$ enclosing the EPs. The solid black line highlights the threshold boundary distinguishing the high-fidelity topological pumping regime from the state mixing regions. Notably, the threshold period scales linearly with the system size $2N$, confirming the high scalability and long-range efficiency of the boundary-driven switching protocol across extended passive bulk. We have chosen $w=1.8v$, and $\theta_0=\pi, s=-1$, and $|\Psi(0)\rangle=[1,0,\cdots]^T$.} \label{fig-5}
\end{figure}

\subsection{Finite-size scaling and encircling time}
\label{sec:scaling}
To evaluate the operational limits of the protocol, we investigate the temporal constraints of the state transfer as a function of the system size $2N$ under a fixed loop radius. Numerically, we observe that while the topological edge states remain strictly isolated within the bulk energy gap for all system sizes, the minimum protocol duration $T_{\text{min}}$ required to achieve non-adiabatic jump for a high-fidelity chiral transfer increases monotonically with $N$ (Fig.~\ref{fig-5}). 

\section{Conclusion and Outlook}\label{sec:conclusion}
In conclusion, we have uncovered a purely boundary-driven mechanism for large-scale non-Hermitian topological chiral pumping across a static bulk in one-dimensional lattices. We have established the precise analytical coordinates for the boundary-induced exceptional points, which remain valid for arbitrarily large system sizes. Furthermore, our protocol demonstrates that dynamically encircling these boundary-localized EPs triggers a robust, direction-dependent chiral switching mechanism, enabling either a high-fidelity cross-chain transfer or a double-switch return of the midgap edge modes, determined by the orientation of the loop topology.

The proposed boundary-controlled topological switch is verifiable  in several state-of-the-art experimental platforms, including active non-Hermitian photonic lattices~\cite{Fritzsche2024,Pan2018,Yu2021},  topolectrical circuits~\cite{Lee2018}, and programmable superconducting qubit arrays~\cite{Erdamar2026, Zhang2025}. In these architectures, the passive SSH bulk is mapped onto standard coupled arrays, while dynamic boundary controls—such as engineered material loss, active operational amplifiers, and microwave dissipation channels, respectively—provide precise real-time tuning of the endpoint potentials. These platforms offer a practical framework for direct observation of the predicted cross-chain state switching and return dynamics.

Looking forward, several directions remain open for future investigation: boundary-driven chiral transfer in higher-dimensional topological phases, such as two-dimensional Chern insulators, or higher-order topological insulators hosting zero-dimensional corner modes; application to open quantum Liouvillian dynamics; and extension to multi-orbital lattices for non-Abelian topological state transfer.\\

{\it Acknowledgments.}~{BM acknowledges financial supports from the ANRF, India (project no.~MTR/2023/000249) and a Seed Grant from IISER Berhampur. }

	\end{document}